\documentclass[aps,prl,twocolumn,floatfix,superscriptaddress,showpacs,amsfonts
,amssymb,amsmath,preprintnumbers]{revtex4-1}
\usepackage{bm}
\usepackage{epsfig}
\usepackage{graphicx}
\usepackage{color}
\usepackage{subfigure}

\newcommand{\be}{\begin{equation}}
\newcommand{\ee}{\end{equation}}
\newcommand{\beq}{\begin{equation}}
\newcommand{\eeq}{\end{equation}}

\newcommand\bea{\begin{eqnarray}}
\newcommand\eea{\end{eqnarray}}

\newcommand{\blue}[1]{\textcolor{blue}{#1}}

\begin{document}
\title{Role of reconnection in inertial kinetic-Alfv\'en turbulence}
\author{Stanislav Boldyrev}
\affiliation{Department of Physics, University of Wisconsin at Madison, Madison, WI 53706, USA}
\affiliation{Space Science Institute, Boulder, Colorado 80301, USA}
\author{Nuno F.\ Loureiro}
\affiliation{Plasma Science and Fusion Center, Massachusetts Institute of Technology, Cambridge MA 02139, USA}

\date{\today}

\begin{abstract}
In a weakly collisional, low-electron-beta plasma, large-scale Alfv\'en  turbulence transforms into inertial kinetic-Alfv\'en turbulence at scales smaller than the ion microscale (gyroscale or inertial scale). We propose that at such kinetic scales, the nonlinear dynamics tends to organize turbulent eddies into thin current sheets, consistent with the existence of two conserved integrals of  the ideal equations, energy and helicity. The formation of strongly anisotropic structures is arrested by the tearing instability that sets a critical aspect ratio  of the eddies at each scale $a$ in the plane perpendicular to the guide field. This aspect ratio is defined by the balance of the eddy turnover rate and the tearing rate, and varies from $(d_e/a)^{1/2}$ to $d_e/a$ depending on the assumed profile of the current sheets. The energy spectrum of the resulting turbulence varies from $k^{-8/3}$ to $k^{-3}$, and the corresponding spectral anisotropy with respect to the strong background magnetic field from $k_z\lesssim k_\perp^{2/3}$ to $k_z\lesssim k_\perp $. 
\end{abstract}

\pacs{52.35.Ra, 52.35.Vd, 52.30.Cv}

\maketitle


\paragraph{Introduction.}
Large-scale low-frequency fluctuations in astrophysical systems such as the interstellar medium, the solar wind, and others, are associated with nearly incompressible magnetohydrodynamic (Alfv\'en) turbulence \cite[e.g.,][]{biskamp2003,elmegreen2004,chen2016}. {The nonlinear Alfv\'en wave packets that compose such turbulence are three-dimensionally anisotropic: elongated along the strong background magnetic field  and resembling current sheets in the perpendicular plane \cite[][]{pouquet1986,biskamp1989,matthaeus_turbulent_1986,biskamp2003,servidio_magnetic_2009,servidio_magnetic_2011,wan2013,zhdankin_etal2013,tobias2013,zhdankin_etal2014,davidson2017,chen2017,boldyrev2005,boldyrev_spectrum_2006,chen_3D_2012,chandran_intermittency_2015,mallet_measures_2016,howes2016,verniero2018}.}  At scales smaller than the plasma microscales, such as the ion gyroscale or ion inertial scale, the shear-Alfv\'en modes transform into kinetic Alfv\'en modes. The character of turbulence then changes qualitatively. Numerical and analytical studies suggest that the energy spectrum becomes relatively steep in the sub-proton range, with the spectral index between $-7/3$ and $-8/3$, and fluctuations become compressible, with density and magnetic field fluctuations comparable to each other~\cite[e.g.,][]{howes_astrophysical_2006,howes08a,howes11a,boldyrev_p12,Groselj2018,roytershteyn2018}. Available solar wind measurements broadly agree with these predictions, with the measured energy spectral slope scattered around a slightly steeper value~$-2.8$ \cite[][]{alexandrova09,kiyani09a,chen10b,chen12a,sahraoui13a}. 

Recently, it has been realized that magnetohydrodynamic (MHD) turbulence becomes affected by the tearing instability at scales larger than the usual Kolmogorov dissipation scale \cite[][]{loureiro2017,mallet2017,boldyrev_2017,comisso2018}, \footnote{Probably, the earliest study of the role of reconnection in MHD turbulence can be found in \cite[][]{carbone1990}. This work, however, was based on the Iroshnikov-Kraichnan model of MHD turbulence \cite[][]{iroshnikov_turbulence_1963,kraichnan_inertial_1965}, which was later understood to be incorrect \cite[e.g.,][]{ng1996,goldreich1997,galtier2000}. A more detailed analysis of the results of \cite{carbone1990} and their comparison to \cite[][]{loureiro2017,mallet2017}, can be found in \cite{boldyrev_loureiro2018}.}. This means that the MHD energy spectrum inevitably gets modified by the tearing instability at small scales. This picture seems to be supported by recent numerical simulations  \cite[][]{walker2018,dong2018}, and it has also been extended to the case of a collisionless plasma \cite[][]{loureiro2017a,mallet2017a,loureiro2018,vech2018}. The role of the tearing instability and reconnection at kinetic scales, however, remains significantly less understood.  Our preliminary dimensional estimates  \cite{loureiro2017a} suggested that the spectrum of tearing-mediated kinetic-Alfv\'en turbulence should become relatively steep, with the spectral exponents ranging from $-8/3$ to $-3$. In the present Letter we propose a self-consistent phenomenological theory of tearing-dominated kinetic-Alfv\'en turbulence.  

Since large-scale Alfv\'en turbulence transforms into kinetic turbulence at small scales, kinetic-scale turbulence is important for energy dissipation in a weakly collisional plasma \cite[e.g.,][]{howes11a,told2015,huang2018}. In addition, kinetic-scale turbulence may be crucial for scattering plasma particles and, in particular, for shaping non-Maxwellian velocity distribution functions of the energetic electrons in the solar wind \cite[e.g.,][]{horaites2018,zank2018}.  Our study also provides a possible theoretical framework for the very recent measurements of electron-only reconnection events in the Earth's magnetosheath~\cite[][]{phan2018} by NASA's MMS mission~\cite{burch_magneto_2016}.  Our results are consistent with these events being related to reconnection-mediated kinetic-Alfv\'en turbulence at subproton scales \cite[][]{loureiro2017a,califano2018}. Indeed, the electron plasma beta is smaller than one in the Earth magnetosheath, meaning that the electron inertial scale, where the magnetic reconnection can operate, is larger than the electron gyroscale, where strong kinetic dissipation comes into play. It is also important to note that a similar ordering of the plasma parameters is also characteristic of the regions close to the solar corona \cite[e.g.,][]{cranmer_etal2009,chandran_etal2011}, which will soon be studied by the recently launched NASA Parker Solar Probe mission \cite[e.g.,][]{bale2016}.  

\paragraph{Model.}
Kinetic-Alfv\'en dynamics retaining electron-inertia effects (the so-called inertial kinetic-Alfv\'en regime) have been investigated recently \cite{chen_boldyrev2017} (see also \cite{passot2017,passot2018}). In the case when the fluctuations are highly oblique with respect to the uniform large-scale magnetic field~${\bf B}_0$, a closed two-field system of equations can be derived that describes the evolution of the magnetic fluctuations $\delta {\bf B}$. In this case $k_z \ll k_\perp$, so one can represent the fluctuating part of the magnetic field as $\delta {\bf B}=-{\hat z}\times \nabla \psi +\delta B_z{\hat z}$. We define the electron skin depth as $d_e=c/\omega_{pe}$, and introduce the dimensionless fields, $\psi'=\psi/(d_e B_0)$ and $b_z=\delta B_z/B_0$, which we will use henceforth, omitting the prime signs for notational simplicity. The resulting system of equations then takes the form~\cite{chen_boldyrev2017}:
\begin{eqnarray}
&\frac{1}{|\Omega_e|}\frac{\partial}{\partial t}\left(1-d_e^2\nabla_\perp^2\right)\psi\nonumber \\
&-d_e^4\left[\left({\hat z}\times\nabla b_z\right)\cdot \nabla\right]\nabla_\perp^2\psi=-d_e\nabla_\|b_z,\label{equation1}\\
&\frac{1}{|\Omega_e|}\frac{\partial}{\partial t}\left(1+\frac{2}{\beta_i}-d_e^2\nabla_\perp^2\right)b_z \nonumber \\
&-d_e^4\left[\left({\hat z}\times\nabla b_z\right)\cdot \nabla\right]\nabla_\perp^2 b_z=d_e^3\nabla_\|\nabla_\perp^2\psi,\label{equation2}
\end{eqnarray}
where {$\beta_i=8\pi n_0 T_i/B^2$} is the ion plasma beta {(assumed order unity)}, and $\Omega_e$ is the electron cyclotron frequency.
The nonlinearities in these equations enter through the terms in the square brackets, and through the field-parallel gradients in the right-hand sides, $\nabla_\|=\partial/\partial z-d_e \left({\hat z}\times \nabla \psi\right) \cdot \nabla $. Since we are interested in the influence of the reconnection effects rather than the dissipation terms on the energy cascade, here we assume that the electron gyroradius is smaller than the electron inertial scale; this corresponds to the electron plasma beta, {$\beta_e=8\pi n_0 T_e/B^2=2\rho^2_e/d^2_e$}, smaller than one. If the small electron gyroradius terms need to be retained, our system~(\ref{equation1}, \ref{equation2}) has to be modified as it is done in \cite[][]{passot2017,passot2018}; such analysis is, however, beyond the scope of the present work as we concentrate only on the scales above the electron gyroradius (and below the ion gyroradius).

Equations~(\ref{equation1}, \ref{equation2}) have two ideal second-order invariants: energy and helicity. At scales large than the electron inertial scale, these invariants have the forms \footnote{For completeness, the full expressions for the energy and generalized helicity, applicable at all scales, are 
$E=\frac{1}{2}\int \left[ b_z \left(1+\frac{2}{\beta_i}-d_e^2\nabla_\perp^2\right)b_z -d_e^2\nabla_\perp^2\psi \left(1-d_e^2\nabla_\perp^2\right)\psi \right] d^3 x,$ and $H=\int \left(1+\frac{2}{\beta_i}-d_e^2\nabla_\perp^2 \right) b_z \cdot \left(1-d_e^2\nabla_\perp^2\right)\psi \, d^3 x .$ 
}: 
\begin{eqnarray}
E=\frac{1}{2}\int\left[\left(1+\frac{2}{\beta_i}\right)b_z^2 +d_e^2 \left(\nabla_\perp \psi \right)^2 \right] d^3\, x,\label{energy}
\end{eqnarray}
and
\begin{eqnarray}
\label{eq:helicity}
H=\int b_z\psi\, d^3\,x.
\end{eqnarray}

\paragraph{Role of helicity.}
The tendency of Kinetic-Alfv\'en turbulence to form current sheets is known from  numerical simulations and observations \cite[e.g.,][]{boldyrev_p12,califano2018,roytershteyn2018,chen2015,sharma2019}. Let us discuss how these may come about.

An analytical description of inertial Kinetic-Alfv\'en turbulence has to account for both the energy and the helicity invariants.
In a turbulent cascade that proceeds from large (MHD) to kinetic scales, the value of the helicity invariant, Eq.~\eqref{eq:helicity}, is set by the turbulence at the ion gyroscale. 
However, in the presence of the energy cascade, helicity cannot cascade to large wavenumbers \cite[e.g.,][]{hasegawa1985,biskamp2003}. 
The selective decay of energy in the presence of conserved helicity can lead to the formation of structures, which can be found from minimizing the functional
\begin{eqnarray}\label{functional}
F=E-\mu H,
\end{eqnarray}
where $\mu$ is the Lagrange multiplier. We thus obtain 
\begin{eqnarray}
& -d_e^2\nabla_\perp^2 \psi=\mu b_z, \label{structures1} \\
& b_z= \mu \left(1+{2}/{\beta_i} \right)^{-1}\psi.\label{structures2}
\end{eqnarray}
{As can be verified, such configurations imply that $E=\mu H$, and the parameter $\mu$ is proportional to the inverse scale of the structure. The configurations minimizing $E$ at constant $H$ correspond to small $\mu$, which reflects he tendency of decaying turbulence to create large-scale helical structures. A similar relaxation to large-scale helical structures is known to exist in magnetohydrodynamic turbulence \cite[e.g.,][]{pouquet1976,biskamp2003}.} 

{One can, however, check that solutions of Eqs.~(\ref{structures1}, \ref{structures2}) correspond to vanishing nonlinearities in Eqs.~(\ref{equation1}, \ref{equation2}). In a turbulent steady-state, however, the nonlinearities cannot vanish as they are responsible for a constant energy flux over scales.
We may nevertheless expect that structures generated in such turbulence should qualitatively resemble those described by Eqs.~(\ref{structures1}, \ref{structures2}); they should lead to reduced, but not identically zero, nonlinear terms.} {Let us denote the characteristic wavenumbers of such a structure in the $x - y$ plane as $k_x$ and $k_y$, which implies $\mu^2\sim k_x^2+k_y^2$. The helicity then scales as $H\sim \mu \psi^2$, while the nonlinear terms are proportional to $k_xk_y \psi^2$. The nonlinear interaction is minimized at constant helicity when one of the wavenumbers becomes significantly smaller than the other, which is consistent with the formation of very anisotropic current sheets. } 

{Let us denote the shorter dimension of such a current sheet as $a\sim 1/k_x$ and the longer one as $l_y\sim 1/k_y \gg a$. A fluctuating magnetic field ${\bf b}_\perp=-{\hat z}\times \nabla \psi$ corresponding to such a structure is aligned with the $y-$direction within a small angle $\sim a/l_y$, while its magnitude  varies on a scale $a$ in the $x-$direction. This, in general, reduces the nonlinear terms in Eqs.~(\ref{equation1}, \ref{equation2}) by a factor $\sim a/l_y$ with respect to their dimensional estimates. For instance, the nonlinear term in the right-hand side of Eq.~(\ref{equation1}) is estimated as $({\hat z}\times \nabla \psi)\cdot \nabla b_z \sim (a/l_y) \psi b_z/a^2$. }

Strongly anisotropic current sheets are thus to be expected in inertial kinetic-Alfv\'en turbulence. 
Motivated by the recent studies mentioned in the Introduction, we may now inquire as to their stability to the tearing mode. 

\paragraph{Tearing instability at sub-proton scales.}
In order to formulate our theory of kinetic-Alfv\'en turbulence mediated by the tearing instability, we first need to analyze the tearing mode arising in a very anisotropic current sheet described by Eq.~(\ref{equation1}, \ref{equation2}). 
We introduce local coordinates such that the $x$ and $y$ axes are across and along the reconnecting current sheet, respectively, and $z$ is along the large-scale field $B_0$. We denote the thickness of the current sheet (in the $x$-direction) as $a$, its length  (in the $y$-direction) as $\sim 1/k_0$, and assume that it is uniform in the $z$ direction. 

We represent the reconnecting part of the magnetic field in such a structure as $\delta {\bf B}_\perp= b_a f(x){\hat y}+{\bf b}_\perp(x,y)$, where $f(x)$ describes the profile of the reconnecting (sheared) field, and ${\bf b}_\perp=-{\hat z}\times \nabla \tilde\psi=\left( \partial \tilde\psi/\partial y, -\partial \tilde\psi/\partial x\right)$ is a small perturbation. 
Regarding the sheared part of the parallel magnetic field, we may assume that it is negligible, so that only the tearing perturbation is left, $b_z=\tilde b_z(x,y)$ \footnote{It can be demonstrated that the presence of sheared component of $\delta B_z$ is {mathematically} analogous to the presence of a sheared velocity field. We have recently shown, however, that it does not qualitatively affect the fastest growing tearing mode~\cite{boldyrev_loureiro2018}.}. 
Following standard procedure, we Fourier transform the perturbations in the $y$ direction:
\begin{eqnarray}
{\tilde \psi}={\bar \psi}(\xi)\exp\left(i k_0 y\right)\exp\left(\gamma t \right),\\
{\tilde b}_z=-i{\bar b}_z(\xi)\exp\left(i k_0 y\right)\exp\left(\gamma t \right), 
\end{eqnarray}
where $\xi=x/a$. 
As we will see, the fastest growing tearing mode is the largest mode that can fit in an eddy, i.e., its lengthscale in the $y$-direction is $\sim 1/k_0$. In what follows, we use only the variables ${\bar \psi}$ and ${\bar b}_z$ and omit the overbar sign. 
Linearizing Eqs.~(\ref{equation1}, \ref{equation2}) with respect to the small perturbations, we obtain the system of equations governing the tearing mode in the kinetic-Alfv\'en case:
\begin{eqnarray}
&\lambda \psi -fb_z=\lambda \frac{d_e^2}{a^2}\psi'',\label{reconnection1}\\
&-f\left(\psi''-\epsilon^2\psi \right)+f''\psi=\lambda b_z''-\lambda \frac{a^2}{d_e^2}\left(1+\frac{2}{\beta_i} \right)b_z,\label{reconnection2}\quad
\end{eqnarray}
where the dimensionless growth rate is $\lambda=\gamma/(k_0 v_{A, e})$, the {\em electron}-Alfv\'en speed is defined with respect to the reconnecting part of the magnetic field, $v_{A, e}=b_a/\sqrt{4\pi m_e n_0}$, and all the derivatives are with respect to~$\xi$. As can be verified after the tearing-mode solution is obtained, $\lambda$ is a small parameter. 
It also turns out that the fastest growing tearing mode corresponds to sufficiently anisotropic current sheets, so that $\epsilon\equiv k_0a$ is a small parameter as well. 

Due to the smallness of the parameter $\lambda$, the right hand sides in Eqs.~(\ref{reconnection1}, \ref{reconnection2}) are relevant only in the inner region of the mode, where the derivatives are large. As will be shown later, the inner scale $\delta_{in}$ is much smaller than $d_e$ in the case of small tearing parameter (the so-called FKR regime), and it approaches $d_e$ in the case of large tearing parameter (the so-called Coppi regime). So, in our analysis we may neglect the last term in the right-hand side of Eq.~(\ref{reconnection2}) as this will not qualitatively change the solution. 

{A useful observation  
is that Eqs. (\ref{reconnection1}, \ref{reconnection2}) describing the kinetic-Alfv\'en tearing instability are then structurally identical to the MHD tearing equations (e.g., Eqs. (11) and (12) in \cite[][]{boldyrev_loureiro2018}) with the only modification that the magnetic diffusivity ${\eta}$ should be replaced by $\gamma d_e^2$ in the kinetic-Alfv\'en case (see a similar discussion in, e.g.,~\cite{bulanov1992,attico2000,zocco_reduced_2011}). } {This reflects the fact that in our case of a collisionless plasma, reconnection is driven by electron inertia rather than resistive effects.} This allows us to straightforwardly derive the growth rate and the corresponding inner scale of the tearing instability. 

The results depend on the assumed profile of the reconnecting magnetic field, $f(\xi)$. 
As in previous work, \cite[e.g.,][]{loureiro2017a,boldyrev_2017,boldyrev_loureiro2018}, we consider two plausible profiles that yield qualitatively different scalings of the tearing mode:  the Harris profile~\cite{harris_1962}, $f(\xi)=\tanh(\xi)$; and a sinusoidal profile, $f(\xi)=\sin(\xi)$.
For the Harris profile, the tearing growth rate in the low $\Delta'$ (FKR) regime is
\begin{eqnarray}
\gamma\sim \frac{v_{A, e}}{a}\frac{1}{(k_0 a)}\left(\frac{d_e}{a}\right)^3,\label{gamma_tanh}
\end{eqnarray} 
and the corresponding inner scale is
\begin{eqnarray}
\delta_{in}\sim d_e \left(\frac{d_e}{a}\right)\frac{1}{(k_0 a)}.
\end{eqnarray}
Both the growth rate and the inner scale increase as the anisotropy parameter $k_0a$ decreases. The anisotropy of the FKR mode cannot, however, be arbitrarily large. The FKR solution is applicable when $k_0a\gg d_e/a$, which justifies our assumption that the inner scale of the tearing mode is smaller than the electron inertial scale.  At $k_0a\sim d_e/a$, the FKR regime crosses over to the Coppi regime (\blue{a large $\Delta'$ regime}), at which point \blue{the tearing mode has the fastest growth rate}
\begin{eqnarray}
\gamma_*\sim \frac{v_{A, e}}{a}\left(\frac{d_e}{a}\right)^2,
\end{eqnarray}
and the inner scales approaches~$d_e$. 

For the sinusoidal magnetic profile, the FKR tearing growth rate is given by  
\begin{eqnarray}
\gamma\sim \frac{v_{A, e}}{a}\frac{1}{(k_0 a)^3}\left(\frac{d_e}{a}\right)^3,\label{gamma_sin}
\end{eqnarray}
and the corresponding inner scale is
\begin{eqnarray}
\delta_{in}\sim d_e \left(\frac{d_e}{a}\right)\frac{1}{(k_0 a)^2}.
\end{eqnarray}
The FKR regime is applicable when $k_0a\gg (d_e/a)^{1/2}$; it transitions to the Coppi regime at $k_0a\sim (d_e/a)^{1/2}$, at which scale the tearing rate reaches its maximal value
\begin{eqnarray}
\gamma_*\sim \frac{v_{A, e}}{a}\left(\frac{d_e}{a}\right)^{3/2}.
\end{eqnarray}
The corresponding inner scale at this point becomes comparable to~$d_e$. 

We now use the tearing growth rates and anisotropies of the corresponding tearing modes just derived to address the influence of the tearing instability on kinetic-Alfv\'en turbulence.

\paragraph{Kinetic-Alfv\'en turbulence mediated by tearing instability.}
The helicity-conservation arguments presented earlier suggest that a typical turbulent eddy in the inertial interval ($d_i \gg a \gg d_e$) is anisotropic in the plane perpendicular to the local mean magnetic field, i.e., $k_0\ll 1/a$.  
From Eqs.~(\ref{equation1}, \ref{equation2}) we may then estimate the characteristic nonlinear time of such a structure as 
\begin{eqnarray}
\gamma_{nl}\sim k_0v_{A, e}(d_e/a).\label{gamma_nl}
\end{eqnarray}  
Obviously, as the anisotropy of the structure increases (that is, $k_0$ decreases), the nonlinear rate~(\ref{gamma_nl}) will decrease, while the tearing growth rate,  Eq.~(\ref{gamma_tanh}) or (\ref{gamma_sin}), will increase. There is, therefore, a limitation on the anisotropy of the structures that can be created by turbulence; very anisotropic structures will be destroyed by the tearing instability. 
By equating the nonlinear rate (\ref{gamma_nl}) to the tearing rate (\ref{gamma_tanh}) or (\ref{gamma_sin}) one finds that, remarkably,  the rates balance when $\gamma=\gamma_*$ in both cases.  
The evolution rate of such a critically anisotropic eddy will, therefore, be governed by the corresponding fastest tearing rate. 

In other words: conservation of helicity subject to energy dissipation suggests the tendency of the turbulence to form elongated, quasi one-dimensional structures. These, however, cannot survive the tearing instability if their field-perpendicular anisotropy is such that the fastest tearing mode fits in such a structure. 
We conjecture, therefore, that the tearing instability is the physical mechanism that sets the eddies' aspect ratio in the field-perpendicular direction.

From Eq.~\eqref{energy} we expect $b_z\sim \nabla_\perp \psi \sim b_a$ at each scale~$a$ in the inertial range of the turbulence. 
We can estimate the energy flux at this scale as $\varepsilon\sim b_a^2\gamma_*={\rm const}$.  We then derive
\begin{eqnarray}    
b_a\propto a d_e^{-2/3},\label{b_tanh} \\
b_a\propto a^{5/6} d_e^{-1/2}\label{b_sin},
\end{eqnarray}
for the $\tanh$ and $\sin$ profiles, respectively. The corresponding Fourier energy spectra of tearing-mediated kinetic-Alfv\'en turbulence are then~\cite{loureiro2017a}
{
\begin{eqnarray}
E(k_\perp) d k_\perp \propto k_\perp^{-3}d k_\perp,\label{energy_tanh}\\
E(k_\perp)d k_\perp \propto k_\perp^{-8/3}d k_\perp.\label{energy_sin}
\end{eqnarray}
}
Our model also allows us to derive the anisotropy of the turbulent eddy with respect to the background magnetic field. The linear frequency of the kinetic-Alfv\'en wave, $\omega=k_zV_{A, e}k_\perp d_e (1+2/\beta_i)^{-1/2}$, where $V_{A, e}$ is a constant electron Alfv\'en velocity defined with the background field~$B_0$. Balancing this equation with the  nonlinear evolution rate $\gamma_*$ at scale $a\sim 1/k_\perp$ (the critical balance condition), we find using Eqs.~(\ref{b_tanh}, \ref{b_sin}) the field-parallel size of the eddy:
\begin{eqnarray}
k_z\propto k_\perp,\\
k_z\propto k_\perp^{2/3},
\end{eqnarray}
for the $\tanh$ and $\sin$ cases, correspondingly. 

The energy spectra (\ref{energy_tanh}) and (\ref{energy_sin}) are not far from those obtained from direct solar-wind measurements, and they seem to be broadly consistent with numerical simulations. Measurements of the guide-field anisotropy are, however, more subtle and they are currently less definitive compared to the measurements of the spectra. Existing phenomenological models, not involving tearing effects, predict energy spectra ranging from $-7/3$ to $-8/3$ and anisotropy relations from $k_z\propto k_\perp^{1/3}$ to $k_z\propto k_\perp^{2/3}$ \cite[e.g.,][]{howes2008,boldyrev_p12,tenbarge2012}; the steepening of the spectrum beyond $-7/3$ is attributed in these models to either the Landau damping or intermittency effects. Our theory, on the contrary, does not invoke dissipation effects or intermittency; the mechanism we propose is different and complementary to the previously developed models. According to our results, steeper energy spectra and weaker $k_z$ {\it vs.} $k_\perp$ anisotropy may be indicative of the presence of tearing effects in kinetic-Alfv\'en turbulence.\\

\paragraph{Discussion and conclusion.}
We have proposed a novel conceptual framework to describe a turbulent cascade in the inertial Kinetic-Alfv\'en regime --- relevant to the Earth's magnetosheath region and to the solar wind in the proximity of the solar corona.
The existence of two ideal invariants of the model equations (energy and helicity) is demonstrated to imply the formation of current sheets, whose instability to the tearing mode becomes a critical physical mechanism governing turbulence at these scales.
As in previous recent work~\cite{boldyrev_2017,loureiro2017a}, we conjecture that the turbulent cascade is determined by the timescale associated with the fastest tearing instability at each scale. {This assumption enables us to compute the expected energy spectra, Eqs.~(\ref{energy_tanh}, \ref{energy_sin}) for two idealized, limiting-case magnetic field configurations: a hyperbolic tangent (Harris) and a sinusoidal profile. Reconnecting magnetic fields occurring in realistic turbulence are probably described by profiles ranging between these two configurations, suggesting that the energy spectra power law exponent may in fact lie between $-8/3$ and $-3$. }

We note that our analysis is also applicable to turbulence of inertial whistlers modes, which generally exist in the same environments as kinetic-Alfv\'en modes but at higher frequencies. Although these modes belong to a different branch of the plasma dispersion relation and exist at frequencies higher than those of kinetic-Alfv\'en modes, the equations governing nonlinear oblique whistler modes are structurally identical to the equations governing the kinetic-Alfv\'en modes. As can be shown \cite[e.g.,][]{chen_boldyrev2017}, the whistler equations can be obtained from system~(\ref{equation1}), (\ref{equation2}) if one neglects the term $1/\beta_i$. 
Our theory is, therefore, applicable to the systems described by electron MHD as well. \\ 

It is worth analyzing our results in the context of MMS' observations of `electron-only' reconnection~\cite{phan2018}.
Note first that any reconnection event occurring in the framework described by Eqs.~(\ref{equation1}-\ref{equation2}) will have a typical outflow velocity of $v_{A,e}$ (the electron Alfv\'en speed computed with the reconnecting component of the magnetic field), consistent with the observations. 
Secondly, we can use our model to compute the scale below which such reconnection cannot involve the ions. Indeed, the length of the current sheet (eddy), set by the tearing mode, is $1/k_0\sim a^2/d_e$ for the Harris profile, or $1/k_0\sim a(a/d_e)^{1/2}$ for a sinusoidal profile. Therefore, requiring $1/k_0 \ll d_i$ results in $ a\ll \sqrt{d_i d_e}$ or $a\ll d_i^{2/3} d_e^{1/3}$, respectively. 
It is encouraging that the measured thickness of the reconnection layer is $a\sim 4d_e$, compatible with these (rough) estimates.

Lastly, we remark that our findings have important implications for numerical studies of turbulence at these scales.
Firstly, the inner scale of the tearing mode occurring in {\it any} eddy in the range $d_i\gg a\gg d_e$ is approximately $d_e$. 
Secondly, helicity conservation (an essential ingredient in the turbulent cascade model that we propose here) implies that correctly capturing this range requires a consistent determination of helicity at the ion scales --- which, in a real plasma, is naturally set by the cascade through the MHD scales. In other words, the usual numerical strategy of studying kinetic-scale turbulence with turbulent forcing imposed at around the ion scales may need to be reconsidered. The implication is that numerical simulations of inertial Kinetic-Alfv\'en turbulence may need to range from large, MHD scales all the way to sub-$d_e$ scales to correctly capture the relevant physical mechanisms at play --- a rather demanding requirement.

\paragraph{Acknowledgments.}
The work of SB was partly supported by the NSF under grant no. NSF PHY-1707272, by NASA under grant no. NASA 80NSSC18K0646, and by DOE grant No. DE-SC0018266. NFL was partially funded by NSF CAREER award no. 1654168 and by the NSF-DOE Partnership in Basic Plasma Science and Engineering, award no. DE-SC0016215.



%

\end{document}